\documentclass[journal=jpccck,manuscript=letter]{achemso}
\SectionNumbersOn

\usepackage[version=3]{mhchem} 
\usepackage{color}
\usepackage{graphics}
\usepackage{amsmath}
\usepackage{nccmath}
\usepackage{epsfig}
\usepackage{fancyhdr}
\usepackage{lineno}
\usepackage{float}
\usepackage[normalem]{ulem}
\usepackage{setspace}
\usepackage{wrapfig}  
\usepackage{soul}
\usepackage{amsmath,amssymb}
\usepackage{tabularx}

\usepackage{url} 
\usepackage{multicol}
\usepackage{multirow}
\usepackage[utf8]{inputenc}
\usepackage[T1]{fontenc}
\usepackage{graphicx}
\usepackage{textcomp}
\DeclareUnicodeCharacter{00E4}{\"a}

\author{Hassan A. Qureshi}
\author{Michael A. Papachatzakis}
\author{Ahmed Gaber Abdelmagid}
\affiliation[Materials UTU]
{Department of Mechanical and Materials Engineering, University of Turku, FI-20014 Turku, Finland}
\author{Mikko Salom\"aki}
\affiliation[Chem UTU]
{Department of Chemistry, University of Turku, FI-20014 Turku, Finland}
\author{Ermei M\"akil\"a}
\affiliation[Phys UTU]
{Department of Physics and Astronomy, University of Turku, FI-20014 Turku, Finland}
\author{Olli Siltanen}
\affiliation[Materials UTU]
{Department of Mechanical and Materials Engineering, University of Turku, FI-20014 Turku, Finland}
\author{Konstantinos S. Daskalakis}
\email{konstantinos.daskalakis@utu.fi}
\affiliation[Materials UTU]
{Department of Mechanical and Materials Engineering, University of Turku, FI-20014 Turku, Finland}

\title{Giant Rabi splitting and polariton photoluminescence in an all solution-deposited dielectric microcavity}

\date{July 2024}

\begin{document}

\section{Abstract}

Planar microcavity polaritons have recently emerged as a promising technology for improving several performance characteristics of organic light-emitting diodes, photodiodes and photovoltaics. 
To form polaritons and achieve enhanced performance, traditional microcavities with high reflectivity mirrors are fabricated by energy-intensive physical vapor deposition methods, which restrict their use in applications requiring flexibility and low cost. 
Here, for the first time, we demonstrate a dielectric all-solution-processed polariton microcavity consisting of Rhodamine 6G films in a poly(vinyl alcohol) matrix, exhibiting more than 400~meV Rabi-splitting and photoluminescence with uniform dispersion along the lower polariton mode. 
Our fully automated deposition and annealing fabrication protocol played a key role in preventing interlayer mixing and producing high optical-quality polariton microcavities, enabling us to observe enhanced scattering of reservoir excitons to the lower polariton and to explore the effects of strong coupling on bimolecular interactions.
Notably, we found that polariton microcavities exhibit a more than 10-fold increase in the critical excitation density for bimolecular annihilation compared to bare Rhodamine 6G films. This enhancement can only be partially attributed to the sub-3-fold measured enhancement in radiative lifetime, highlighting the critical role of strong coupling in the influence of molecular dynamics.

\section{Introduction}
Engineering device architectures that facilitate strong interactions between absorbers and their electromagnetic environment is central to modern organic optoelectronics. When the energy exchange rate between a material and its environment surpasses their intrinsic losses, new hybrid light-matter states form called polaritons\cite{Kavokin2011,torma_strong_2015,FriskKockum2019}. 
In recent years, polaritons in molecular systems have emerged as a promising platform for modifying chemical reaction rates\cite{Feist2018, Garcia-Vidal2021, Thomas2023, Dutta2024}, emission kinetics\cite{coles2014polariton,ribeiro2018polariton,Berghuis2019,Polak2020,Yu2021,Abdelmagid2024,Ishii2024}, exciton diffusion\cite{Bhuyan2023, Xu2023, Pandya2021, Balasubrahmaniyam2023}, and for enhancing the performance characteristics of organic light-emitting diodes\cite{Graf2017,genco2018bright, Daskalakis2019, mischok2023, zhao2024stable, de2024organic}, photodiodes\cite{ eizner2018organic, mischok2020spectroscopic, Wang2021}, and photovoltaics\cite{Nikolis2019, Sokolovskii2023}, 
as well as introducing novel coherent light sources and all-optical switches\cite{Daskalakis2014, Plumhof2013,  Lerario2017, Rajendran2019, Zasedatelev2021, Tang2021, Moilanen2021, jiang2022exciton}. In essence, these effects originate from the hybrid nature of polaritons, which inherit properties from both light and matter: light contributes to delocalization, while excitons provide strong interaction terms. Planar microcavities are a versatile photonic architecture for modifying the electromagnetic environment; they are easy to design, fabricate, and study under optical and electrical excitation. Thus, they have been extensively used in polariton studies in mirror configurations of metal-metal, dielectric-dielectric or a combination of both\cite{palo2023}. 

Typical metal-metal (or metal-clad) polariton microcavities use aluminum (Al), silver (Ag) or gold (Au) to achieve high reflectivity in the ultraviolet, visible and infrared ranges, respectively. In these designs, a highly reflective, sub-100-nm-thick, bottom mirror and a semitransparent, sub-30-nm-thick, top mirror sandwich the active material. This scheme allows for minimal physical deposition runs and reduced mode volume---important for increasing the exciton-cavity interaction which scales as $\sim \sqrt{N/V}$, where $N$ is the number of absorbers and $V$ is the mode volume \cite{Kena-Cohen2013}. However, the semitransparent top mirror is a bottleneck for real-world applications of microcavity polaritons. It blocks light from going in and absorbs the light the dipoles emit via single-plasmon and surface-plasmon resonances. Importantly, because of its lossy nature, it also limits the quality factor (Q) of such microcavities to below 90. This is detrimental to enhancing the radiative rates of the cavity emitter---the Purcell enhancement factor being proportional to $Q/V$---which is one of the main reasons for using microcavities in the first place.  

\begin{figure}
\vspace{-40pt}
\centering
\includegraphics[width=\linewidth]{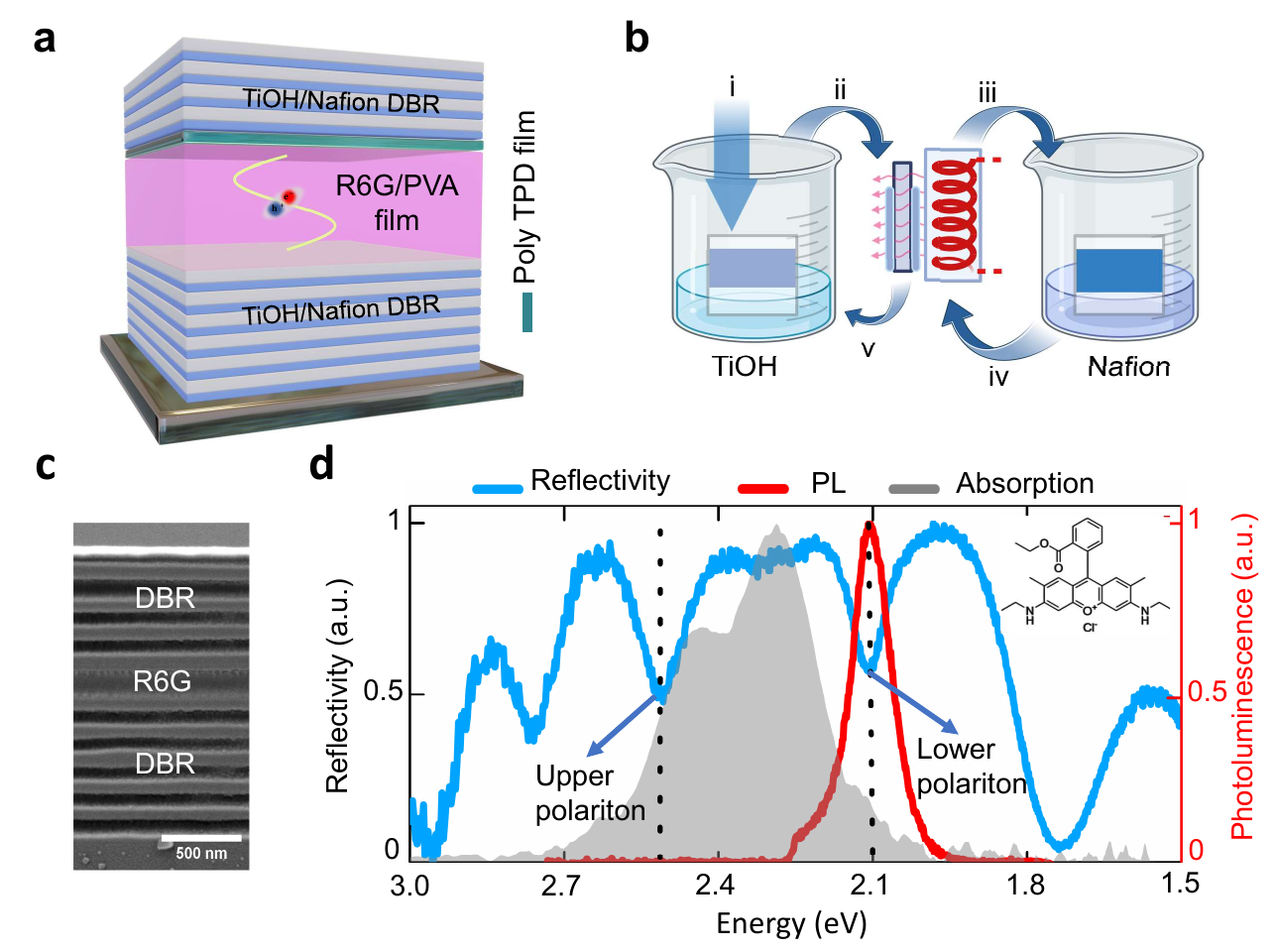}
\vspace{-20pt}
\caption{(a) Schematic showing the fabricated all-solution-based microcavity consisting of alternating layers of Nafion and titanium hydroxide / polyvinyl alcohol hybrid films. Optical measurements were performed on samples with 6-pair bottom and 5-pair top DBRs. A 20-nm-thick poly(N,N'-bis-4-butylphenyl-N,N'-bisphenyl)benzidine buffer film spin-coated on Rhodamine 6G and Ossila E132 encapsulation epoxy was carefully dip-coated at the edges of the half cavity to protect Rhodamine 6G from dissolving when depositing the second DBR. The combination of poly(N,N'-bis-4-butylphenyl-N,N'-bisphenyl)benzidine buffer film and epoxy edge sealing acts as our scheme for encapsulation of the half-cavity. (b) Illustration of the automatized dip-coating setup for the fabrication of DBRs. For every DBR pair we repeat the following process: \romannumeral 1) dipcoating the titanium hydroxide / poly(vinyl alcohol) solution, \romannumeral 2) annealing, \romannumeral 3) dipcoating Nafion, \romannumeral 4) annealing, and \romannumeral 5) repeating the process until the desired number of pairs is deposited. (c) Typical scanning electron microscopy cross section of a full microcavity. The image is taken from 5-bottom, 4-top pair microcavity that was etched using a broad ion beam. (d) Normal incidence reflectivity spectrum (blue line, left axis) of a microcavity with 3 mg/ml concentration resulting in an $\Omega_R$ of 264 meV and photoluminescence at normal incidence from the lower polariton (red line, right axis). The absorption spectrum of a bare 70-nm-thick Rhodamine 6G film with 3 mg/ml, as used in the microcavity, is shown as greyed-out area.}
\label{fig:1}
\vspace{0pt}
\end{figure}

A better alternative is dielectric distributed Bragg reflectors (DBRs). They are periodic structures consisting of multiple layers of alternating materials with varying refractive indices. They can be engineered to reflect only a specific range of wavelengths while remaining transparent outside that range. Their reflectivity can be easily increased by adding more layer pairs (with the expense of narrowing the reflectivity stopband), enabling the fabrication of microcavities with exceptionally high quality factors Q \cite{zhangfabrication, bronnbauer2018printing, Bachevillier2019, palo2023}. DBRs have negligible absorption and plasmon-related losses, which is why they are commonly used in polariton lasing experiments. However, DBR deposition is typically achieved by using sophisticated, resource and time-demanding deposition processes such as physical-vapour deposition (PVD), atomic layer deposition (ALD), and molecular beam epitaxy (MBE)\cite{lova2018advances}. In most cases, these deposition methods are unfit for organic optoelectronics where they consist of sensitive organic layers. In fact, until recently, organic polartion DBR microcavities have been exclusively made with PVD\cite{palo2023developing, Strang2024}.
  

Here we report, for the first time, polariton photoluminescence dynamics from a dielectric DBR planar microcavity fabricated entirely using solution-processing methods, specifically dip-coating and spin-coating. This fabrication process holds the potential to significantly reduce costs while maintaining or even exceeding the performance of conventional techniques. Our primary focus is on developing a robust and versatile fabrication process, for which we selected Rhodamine 6G (R6G) as the active material due to its extensive characterization in the literature. 
As confirmed by direct comparisons with results documented in the literature\cite{Hakala2009,Hulkko2021}, our fabrication process consistently produces pristine polariton microcavity samples. With this approach we ensure that our methods prevent interlayer mixing and maintain the integrity of the active layer. 
Notably, our microcavities demonstrated Rabi splitting ($\Omega_R$) larger than 400~meV, matching the one observed in metal-clad microcavities filled with identical R6G concentrations~\cite{Tanyi2017,hu2020solution} (Supplementary Fig.~\ref{meta_clad}), and surpassing by 10-fold the coupling strengths of previous all-solution-based microcavities containing a perylene diimide derivative\cite{Strang2024}. 

We support our results with angle-resolved reflectivity and photoluminescence measurements,
which we compare to PVD-fabricated silver-clad microcavities (Supplementary Fig.~\ref{meta_clad}). Additionally, excitation-density-dependent measurements reveal that in polariton microcavities, singlet-singlet annihilation (SSA) is significantly suppressed compared to bare R6G films. 
Time-resolved photoluminescence measurements show a sub-3-fold reduction in radiative lifetime, which can be attributed to enhanced scattering of reservoir excitons to the lower polariton. This reduced singlet depopulation lifetime partially explains the observed SSA suppression, highlighting the rich physics of strong coupling.


\section{Results}
Figure \ref{fig:1}a shows a schematic representation of the developed microcavities. They consist of an asymmetrical number of DBR layer pairs, 6 for the bottom and 5 for the top mirrors. The DBRs are fabricated using an automatized sequential dip-coating and annealing protocol, reported in our previous work\cite{palo2023developing}, of Nafion (low refractive index) and titanium hydroxide / poly(vinyl alcohol) (TiOH/PVA as high refractive index) films (see Fig.~\ref{fig:1}b and Methods). The R6G molecules were embedded in a polyvinyl alcohol (PVA) matrix and spin-coated on the bottom DBR. To prevent the R6G/PVA active materials from dissolving during the deposition of the top DBR, a 20-nm-thick poly(N,N'-bis-4-butylphenyl-N,N'-bisphenyl)benzidine (poly-TPD) transparent film was spin-coated on the R6G/PVA film. We found that our glass substrates had micrometer-sized "teeth-like" roughness along their edges, through which we observed detrimental cross-diffusion between R6G/PVA films and DBR solutions (see Supplementary Fig.~\ref{UV_glue_and_Substrate}). To prevent this, we carefully dip-coated the substrate edges with a UV-curing epoxy (see Methods and Supplementary Information). 

We wish to emphasize that the encapsulation steps using poly-TPD and UV-curing epoxy were crucial for producing high-quality solution-based polariton microcavities. In solution-based coating methods, it is common to focus on solvent orthogonality between layer depositions to prevent cross-mixing. However, for solution-based DBRs to become a viable technology for polaritonics, it is essential to develop a standardized DBR architecture that can accommodate any active material or concentration, providing fabrication flexibility comparable to PVD. Our approach is both simple and effective, and it can be easily extended to active materials beyond R6G. Figure \ref{fig:1}c shows a cross-sectional scanning electron microscopy image of our typical samples. It confirms the consistency of the polariton microcavity layers and that cross-dissolvement was prevented. Note that some fragmentation in the intracavity layers is likely due to the broad ion-milling parameters used to produce the crossectional images. Moreover, nm-sized blobs are visible in the TiOH/PVA DBR layers which can be either TiO$_2$ nanoparticles from incomplete hydrolysis or hotspots generated after milling. In our previous work\cite{palo2023developing}, ellipsometry measurements of the TiOH/PVA did not show a scattering effect in the visible spectral range, further confirming that TiO$_2$ nanoparticles are either totally absent, or their size is sub-50-nm. 

\begin{figure}
\vspace{-5pt}
\centering
\includegraphics[width=\linewidth]{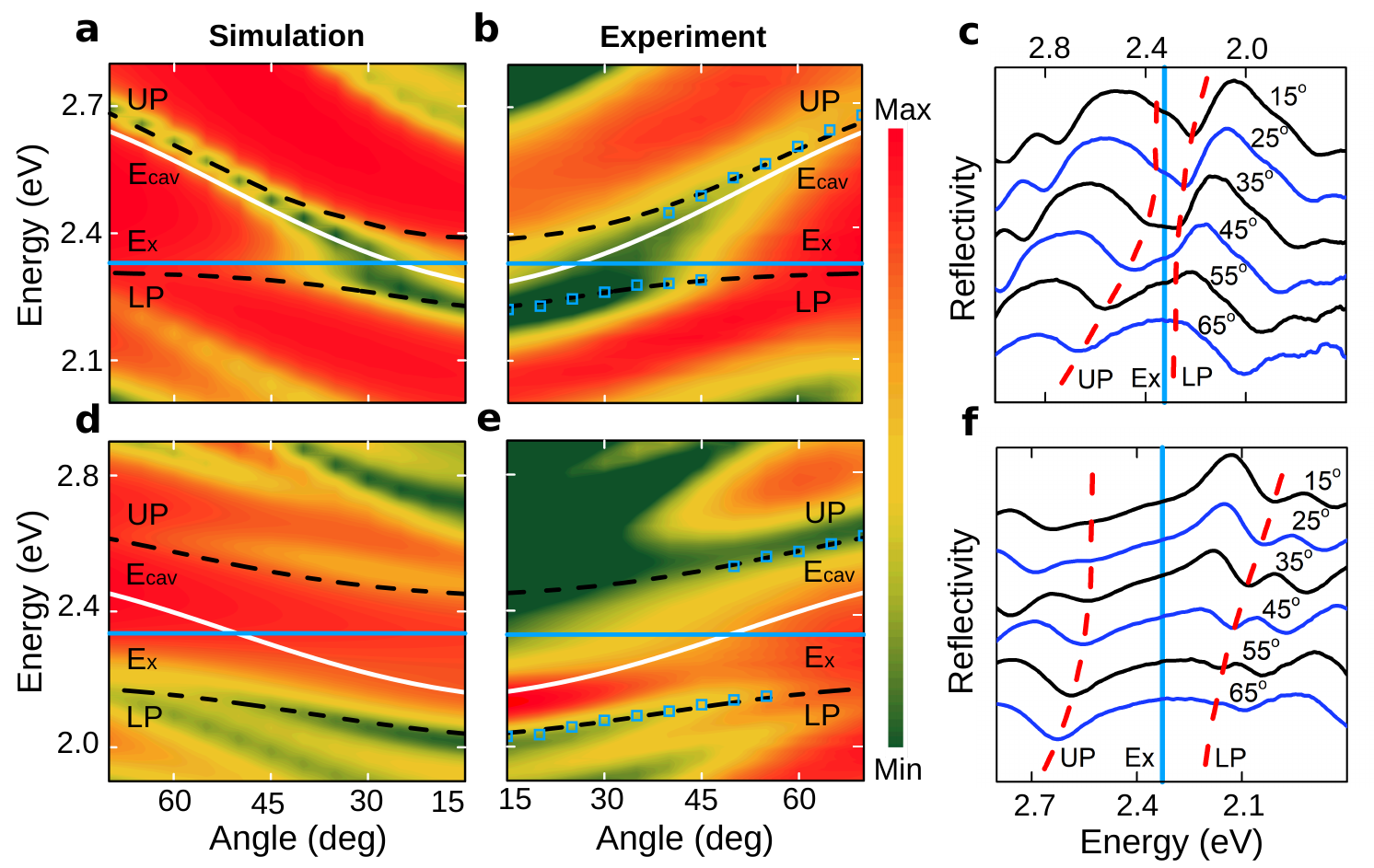}
\vspace{-12pt}
\caption{Contour maps of transverse electric (a),(d) transfer matrix simulations and (b),(e) angle-resolved reflectivity measurements for the 1 mg/ml and 5 mg/ml R6G DBR microcavities. In Supplementary Fig. \ref{Reflectivity_Moderate_Conc} we show the plots for the 5 mg/ml R6G DBR microcavity. In (b),(e), the positions of individual minima identified on the angle-resolved spectra (c),(f) are shown as blue squares. The dashed lines are a fit of the lower polariton (LP) and upper polariton (UP) minima to a coupled harmonic oscillator model while the solid lines show the uncoupled cavity photon (E$_{cav}$) and exciton (E$_X$) dispersion relations}.
\label{fig:2}
\vspace{-5pt}
\end{figure}

To examine the dependence of coupling strength for increasing R6G concentration, we fabricated three polariton microcavities with concentrations of R6G in a PVA matrix of 1~mg/ml, 3~mg/ml and 5~mg/ml, resulting in $\Omega_R$ of 130~meV, 264~meV and 401~meV, respectively. $\Omega_R$ was extracted by fitting a coupled harmonic oscillator in the measured reflectivity spectra (Fig.~\ref{fig:1} and Supplementary Fig.~\ref{Reflectivity_Moderate_Conc}). The microcavities were precisely designed and fabricated to align the optical mode to the R6G J-aggregate absorption peak\cite{Tanyi2017}, shown in Supplementary Fig.~\ref{R6G}. Thus, in all three polariton microcavities, the active layer thickness was kept at around 73~nm (see Methods for fabrication and characterization details). Figure~\ref{fig:1}d shows normal incidence reflectivity from the polariton microcavity with 3~mg/ml R6G concentration. It demonstrates clearly that upper and lower polariton modes coexist within the DBR stopband, together with Bragg modes at the edges of the stopband. For completeness, we also plotted the photoluminescence from the lower polariton mode, which is discussed in detail later in this paper.

Figure~\ref{fig:2} shows the reflectivity from polariton microcavities with 1 and 5 mg/ml R6G concentrations, respectively. Supplementary Fig.~\ref{Reflectivity_Moderate_Conc} shows reflectivity spectra from the microcavity with 3 mg/ml R6G concentration. Increasing the concentration of R6G in the PVA matrix slightly increases the effective refractive index and the actual thickness of the spin-coated films; the 5 mg/ml films are usually around 3~nm thicker than 1 mg/ml for same spin-coating parameters. Therefore, the microcavity with high concentration exhibits a slightly higher effective cavity thickness than the low concentration microcavity, resulting in a small redshift of the uncoupled cavity $-0.110$ meV. 

\newcommand\varpm{\mathbin{\vcenter{\hbox{%
  \oalign{\hfil$\scriptstyle+$\hfil\cr
          \noalign{\kern-.3ex}
          $\scriptscriptstyle({-})$\cr}%
}}}}

Despite that, the main origin of the large redshift of the lower polariton is due to the substantially increased coupling strength. Namely, the hybridized eigenenergies are given by~\cite{Bhuyan2023}
\begin{equation}
    E_{UP(LP)}=\frac{E_X+E_{cav}}{2}\varpm\sqrt{\Big(\frac{\Omega_R}{2}\Big)^2+\Big(\frac{E_X-E_{cav}}{2}\Big)^2},
    \label{eq:eigenenergies}
\end{equation}
where $E_X$ and $E_{cav}$ are the singlet and cavity mode energies, respectively.

As expected, increasing the coupling strength leads to a greater splitting between the upper and lower polariton modes, causing the lower polariton to flatten and redshift with respect to the uncoupled cavity resonance. 
Figures \ref{fig:2}c,f provide angle-resolved reflectivity spectra for better visibility of the polariton dips. Additionally, Figures \ref{fig:2}a,d present transfer matrix reflectivity maps of the designed samples, showing an excellent agreement with the experimental results.

To benchmark the performance of our solution-processed polariton microcavities, we fabricated silver-clad microcavities using the same R6G concentrations as those in the DBR microcavities. The silver mirrors were fabricated using ultra-high vacuum thermal evaporation, and the R6G/PVA films were spin-coated using the same recipe. As shown in Supplementary Fig.~\ref{meta_clad}, our DBR microcavities matched the coupling strength of the Ag-clad microcavities. Generally, metallic-clad microcavities exhibit smaller mode volumes than their DBR counterparts which suffer from effective cavity thickness losses. However, the low Q-factor of metallic-clad microcavities leads to reduced interaction strength. In contrast, our DBR microcavities achieve a higher Q of more than 90\cite{palo2023developing}, with a minimal number of DBR pairs, which plays a crucial role in achieving higher coupling strengths. To minimize potential damage to the R6G/PVA active layer during the deposition of the top silver mirror, we used low deposition rates, which have been previously reported to preserve sample integrity\cite{Hulkko2021}.

Note that for the R6G concentration range studies in this paper, we see that R6G absorption is dominated by H- and J-aggregation, and both contribute in the coupling\cite{Tanyi2017,Hestand2018}. In Supplementary Fig.~\ref{R6G} we show the absorption measurements of 70~nm R6G in PVA films having the same concentration as to those used in the microcavity samples namely 1 mg/ml, 3 mg/ml and 5 mg/ml. We observe that J-aggregate has a stronger contribution for all studied concentrations. H-aggregate contribution is still present at all concentrations, but its contribution becomes more significant with higher R6G concentrations. The contribution of the two species of aggregates is also apparent in the photoluminescence from the bare films where we see that emission is $-172$ meV shifted compared to the low concentration. Despite this behavior of thin films, we found that $\Omega_R$ still increases linearly with concentration (Supplementary Fig.~\ref{Rabi vs concentration}). 

\begin{figure}
\vspace{-20pt}
\centering
\includegraphics[width=\linewidth]{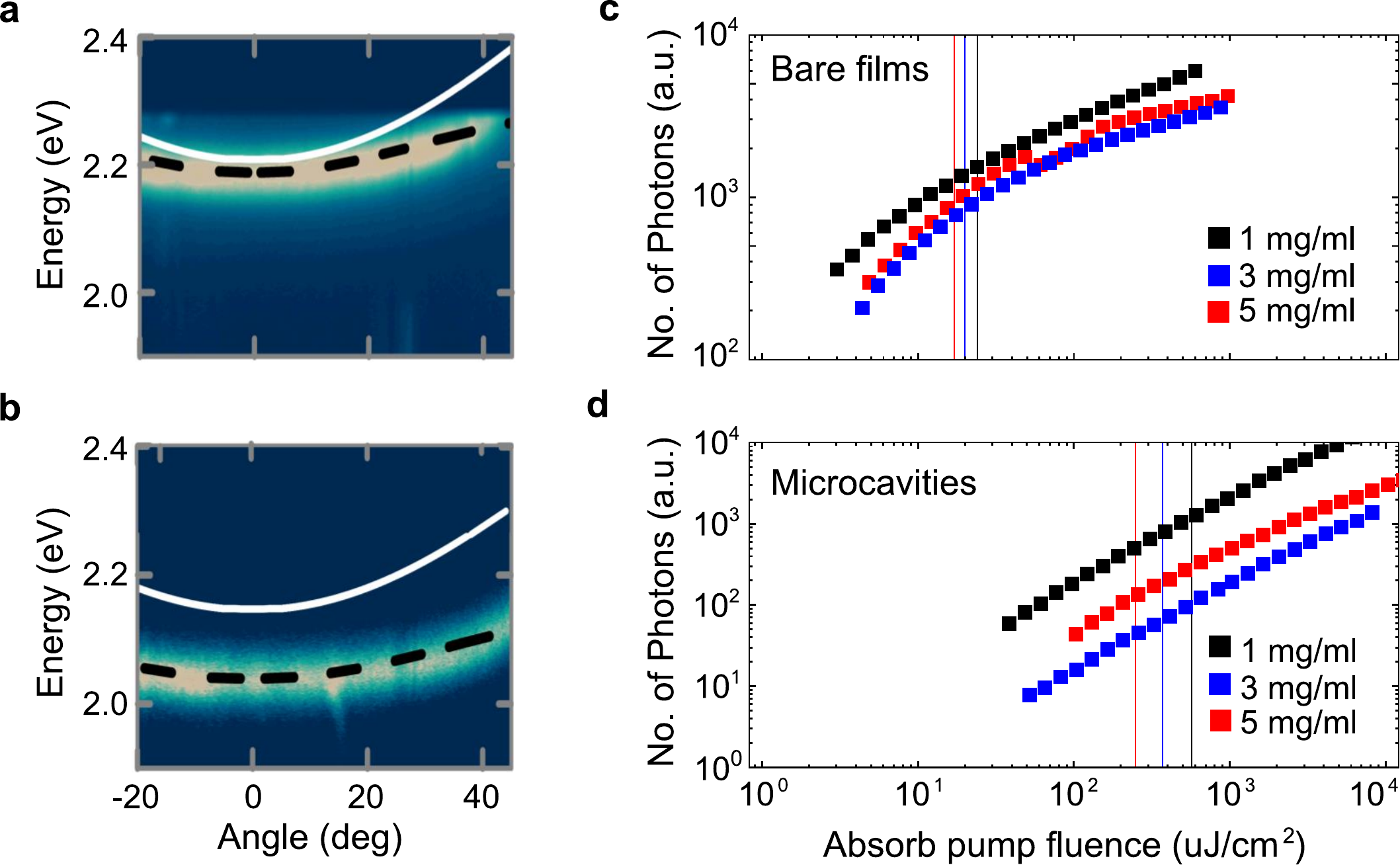}
\vspace{-12pt}
\caption{(a),(b) Angle-resolved photoluminescence at low pump fluence ($\sim$ 50 \textmu J/cm$^2$) for microcavities with 1 mg/ml and 5 mg/ml R6G concentrations. The microcavity with high concentration exhibits a slightly higher effective cavity thickness than the low concentration microcavity, resulting in a small shifting of $-0.172$ meV. Despite that, the main origin of the large redshift of the lower polariton is due to the substantially increased coupling strength. (c),(d) The dependence of the photoluminescence photon number on the absorbed pump fluence is shown for the three different R6G concentrations for (c) bare films and (d) microcavities. SSA is assumed to start dominating at the value of absorbed pump fluence at which it deviates 10~\% from the initial linear trend. For bare films, significant SSA was observed at approximately 17, 20, and 24~\textmu J/cm$^2$ in descending order of concentrations. The corresponding pump fluences for microcavities were 249, 371, and 570~\textmu J/cm$^2$. These values are indicated by the vertical lines in (c) and (d). 
}
\label{fig:3}
\vspace{0pt}
\end{figure}

Under optical excitation, all polariton microcavities exhibited uniform-intensity photoluminescence dispersion along the lower polariton mode, as shown in Figure \ref{fig:3}, where angle-resolved photoluminescence is plotted for 1 mg/ml and 5 mg/ml concentrations (see Supplementary Fig.~\ref{Moderate_polariton_PL}). The black dashed lines represent the dispersion of the lower polariton, extracted from reflectivity measurements, which perfectly matches the emission dispersion. The white line shows the uncoupled microcavity dispersion obtained from coupled harmonic oscillator calculations. We observed that photoluminescence maintained uniform intensity across the entire angular dispersion for all three samples which further confirms strong coupling. Uniform emission dispersion can be attributed to efficient radiative pumping\cite{Hulkko2021}. As shown in Supplementary Fig.~\ref{emisison_boradenning}, both the bare films and microcavities exhibit the highest photoluminescence quantum yield at a concentration of 1 mg/ml. Generally, photoluminescence intensity is decreased with increasing concentration due to aggregation\cite{Chandrasekhar1943}. Surprisingly, the 3 mg/ml samples show the lowest photoluminescence quantum yield, which we speculate is due to competition between J- and H-aggregate emission\cite{Tanyi2017}. 

\begin{figure}
\vspace{-5pt}
\centering
\includegraphics[width=\linewidth]{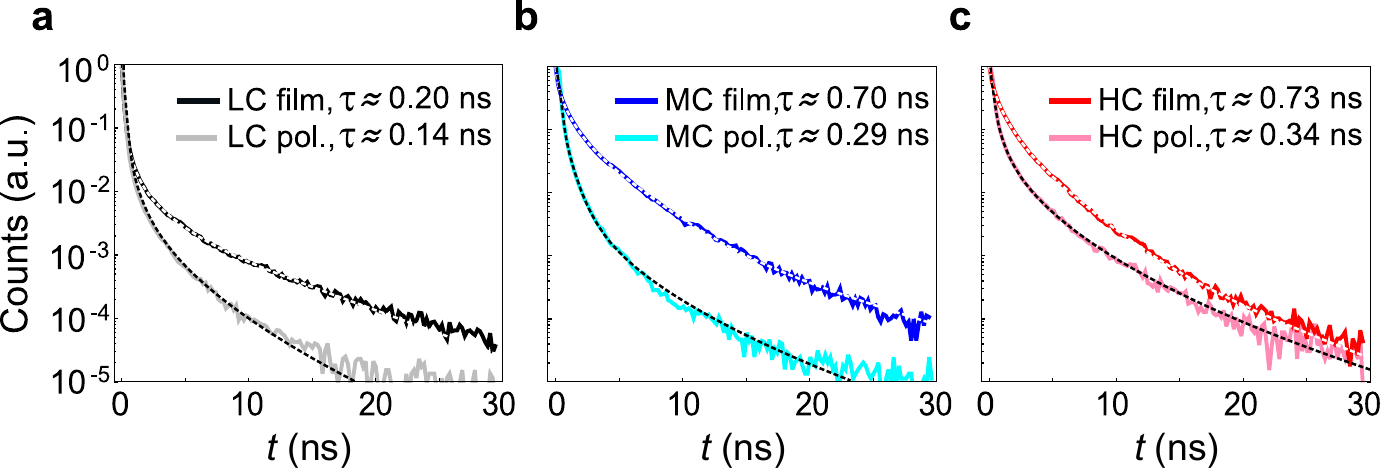}
\vspace{-12pt}
\caption{Transient photoluminescence measurements and lifetime fittings. (a) Low-concentration (LC, 1 mg/ml) film and polariton. (b) Medium-concentration (MC, 3 mg/ml) film and polariton. (a) High-concentration (HC, 5 mg/ml) film and polariton. The dashed curves are the fitted functions.}
\label{fig:4}
\vspace{0pt}
\end{figure}

Upon increasing the excitation intensity, the bare 70-nm-thick R6G films exhibited a sublinear increase in photoluminescence (Fig.~\ref{fig:3}c and Supplementary Fig.~\ref{emisison_boradenning}). We can reasonably conclude that the biomolecular annihilation observed is predominantly due to singlet-singlet annihilation (SSA), as exciton-plasmon interactions are ruled out by the use of non-metallic mirrors, and the triplet population is negligible given that the singlet-triplet energy gap in Rhodamine 6G exceeds 1~eV\cite{Kordel2010}. These bare films are essentially identical to those used in the corresponding polariton microcavities. Naturally, the severity of SSA was greater for the film with the highest concentration. In particular, we defined the critical absorbed pump fluence as the value at which it deviates 10~\% from the initial, linear trend on the log-log scale (cf. Ref.~\cite{Murawski2013}). After this value, SSA and other quenching mechanisms start to dominate exciton dynamics. For bare films, the critical absorbed pump fluences are 24, 20, and 17~\textmu J/cm$^2$ for 1 mg/ml, 3 mg/ml and 5 mg/ml respectively. In contrast, for polariton microcavities the critical absorbed pump fluences are 570, 371, and 249~\textmu J/cm$^2$ for 1 mg/ml, 3 mg/ml and 5 mg/ml respectively (Fig.~\ref{fig:3}d and Supplementary Fig.~\ref{emisison_boradenning}).

To investigate the origin of the SSA suppression, we performed transient photoluminescence measurements on bare films and polaritons at low pump excitations (Fig.~\ref{fig:4} and Methods). The polariton microcavities showed a reduction in radiative lifetime $\tau$ by less than 3-fold compared to bare R6G films. This reduction is attributed to faster depopulation of the exciton reservoir, which raises the critical excitation density at which SSA occurs (see \cite{zhao2024stable} and reference therein). This radiative enhancement alone accounts for part of the suppression of SSA. Assuming that the critical pumping rate scales as $\tau ^{-2}$~\cite{Baldo2000}, this would lead to a sub-10-fold increase in excitation density required for SSA onset. However, in all polariton microcavities, we observed more than a 10-fold increase in the SSA threshold, namely 24-, 19-, and 15-fold for 1 mg/ml, 3 mg/ml and 5 mg/ml respectively. In addition, radiative lifetime of bare films could be fitted with simple exponentials. In contrast, polariton microcavity lifetimes required stretched exponential fits (see Methods). All of the above hints that polariton emission in our samples results from competing physical processes. Although a comprehensive study of this novel physical mechanism is planned, we speculate two possible scenarios that could explain it. The combination of large coupling strength and large photonic content of 64~$\%$ (Supplementary Fig.~\ref{Hopfield}) reduces the exciton effective mass allowing them to propagate away from neighboring excitons ballistically~\cite{Lerario2017,Xu2023,Pandya2021,Balasubrahmaniyam2023}, effectively reducing the SSA capture radius and thus lowering the probability of SSA\cite{Chandrasekhar1943, Wallikewitz2012}. The second scenario is based on our recent theoretical study, which predicts that under sufficiently strong coupling, singlets couple more efficiently to the cavity mode than they interact with neighbouring excitons, leading to reduced SSA~\cite{siltanen2024incoherentpolaritondynamicsnonlinearities}. This is in contrast to molecules with strong singlet fission, where strong coupling typically enhances singlet fission~\cite{Martinez-Martinez2018,Climent2022}.

\section{Summary}
This study presents the development of a fully solution-processed dielectric microcavity incorporating Rhodamine 6G (R6G) as the active material, achieving a substantial Rabi splitting of over 400~meV while also demonstrating photoluminescence from the lower polariton mode. The microcavities were fabricated using spin-coating and an automated dip-coating process, optimized to prevent interlayer mixing and maintain the integrity of the active layers. Reflectivity and photoluminescence measurements confirmed the pristine quality of the samples, with coupling strengths that not only matched but, in some instances, surpassed those of metallic-clad microcavities. The study also investigated the excitation dependence of photoluminescence. At low pump fluences, all microcavities displayed uniform photoluminescence intensity, with intensity decreasing as R6G concentration increased due to aggregation. Notably, polariton microcavities exhibited suppressed SSA compared to bare R6G films, highlighting the critical role of coupling strength in shaping photophysical behaviour. This work demonstrates the viability of solution-based fabrication techniques for creating advanced polaritonic devices, with broad implications for improving the performance of organic optoelectronic systems and potentially addressing one of the longest-standing challenges in solid-state organic laser diodes—efficiency roll-off caused by exciton-exciton annihilation\cite{Giebink2008,Yoshida2023}.

\section{Methods}
\subsection{Materials and Thin film fabrication}
The high refractive index titanium hydroxide / polyvinyl alcohol hybrid films were synthesized using titanium (IV) chloride (Quality level 100) and PVA (MW 30000-70000) both purchased from Sigma Aldrich. The stock solution of TiOH was made by slowly hydrolyzing 2.2 ml of TiCl4 in 20 ml of cold deionized H20 in an ice bath. This solution was then added to aqueous PVA in 1:1 volume to create the TiOH/PVA hybrid solution as described by \cite{Russo2012}. Titanium hydroxide/ PVA hybrid films were fabricated using an in-house modified Ossila dip coater with a single-step deposition and annealing. 
To fabricate the low refractive index films, we diluted Nafion into desired concentrations in IPA and then deposited it using the same dip coater. Nafion D520 (5\% dispersion in water and low aliphatic alcohols) was purchased from Chemours while IPA was purchased from Sigma-Aldrich.
 Rhodamine 6G was purchased from Luminescence Technology. R6G stock solutions were synthesized by adding methanol in a desired amount of R6G. The stock solutions were then combined 1:1 in volume with aqueous PVA to make an R6G/PVA solution. The solutions were spin-coated and annealed to fabricate R6G/PVA active films.
\subsection{Sample fabrication}
The substrates (15mm*15mm*1mm) were cleaned with water/soap (3 \% Decon 90) and Isopropanol solutions to remove any residue from the surfaces. The substrates were sonicated for 10 minutes on each step and finally dried with a nitrogen purge. A similar cleaning protocol was deployed for the silicon substrates that were used to measure the optical constants and thicknesses with ellipsometry\cite{Leppala2024}. 
As illustrated in Fig.~\ref{fig:1}a), microcavities consisting of two DBRs separated by an active layer were fabricated on glass substrates. In each deposition step, the substrate was lowered into the desired solution and retracted at a constant speed of 40 mm/min to ensure uniform surface wetting. The coated substrate was then placed on a heating element and dried at 80 °C for one minute. Before proceeding to the next layer, the sample was dried in the ambient temperature for one minute to dissipate heat. The process is repeated until the desired number of depositions was achieved, resulting in a DBR structure on both sides of the substrates. We removed the DBR facing the heater by scraping it out. The 70 nm active material film was spin-coated from the R6G/PVA solution and dried at 80 °C for two minutes. The 20 nm poly-TPD protective film was spin-coated and dried at 100 °C for 10 minutes. To protect the edges, Ossila E132 encapsulation epoxy was carefully dip-coated at the edges. The sample was soft-baked at 100 ° C for two minutes, exposed to UV for five minutes, and then the top DBR was deposited with the same dip-coating process. 

\subsection{Optical Characterization}
The optical constants and thicknesses of the thin films were acquired using a J.A Woollam VASE ellipsometer. We utilized an Xe lamp with a spectral range of 250-2500 nm to obtain the spectra, the data was analyzed by fitting a Cauchy model in the transparent region of the film. The dispersion of the DBRs and polaritonic modes was obtained with a VASE ellipsometer and a custom-built angle-resolved imaging setup that can measure reflectivity and photoluminescence. The setup uses a collimated light from a halogen lamp illuminating the sample through a 0.75 NA microscope objective. The reflected light is then collected with the same objective and the back focal plane image is then focused into the slit of the spectrometer that is coupled to a two-dimensional (2D) CCD camera (1340*400 pixels). The reflectivity dispersion was then resolved in wavelength vs angle. The sample was excited in PL configuration using 250 fs pulses at 530 nm and 200 kHz repetition rate (Light conversion Pharos, Orpheus and Lyra). The TMM was used to simulate the reflectivity of DBR and a couple harmonic oscillator model was used to fit the polariton modes and extract the Rabi splitting energy.

\subsection{Fitting}
The lifetimes $\tau$ in Figure~\ref{fig:4} were evaluated by fitting a generic multi-exponential model
\begin{equation}
    I(t)=\sum_iA_ie^{-(t/\tau_i)^{\beta_i}},\hspace{5pt}\sum_iA_i=1
    \label{eq:multiexp}
\end{equation}
to each (normalized) data set and calculating~\cite{Ye2021}
\begin{equation}
    \tau=\int_0^\infty dtI(t).
    \label{eq:lifetime}
\end{equation}
Here, $A_i$ is the weight of the decay channel $i$ with the decay time $\tau_i$. The exponents $\beta_i$ either stretch ($0<\beta_i<1$) or compress $(1<\beta_i)$ the functions. In the bare-film cases, it was enough to fix $\beta_i=1$. However, this did not give us good fits in the polariton cases, which is why we set the parameters $\beta_i$ free. 

\subsection*{Author Contributions}
KSD conceived the project, designed the structures and guided the experiments. HAQ performed all the fabrications and characterization of the samples. MP built the automatized fabrication DBR system. AGA fabricated metal-clad microcavities and helped in insulating the DBR samples. OS performed the time-resolved photoluminescence analysis. HAQ, OS and KSD wrote the manuscript. All authors contributed to the draft, discussion, and analysis of the data.

\subsection*{Conflicts of interest}
There are no conflicts to declare.
\newline


\begin{acknowledgement}
This project has received funding from the European Research Council (ERC) under the European Union’s Horizon 2020 research and innovation programme (grant agreement No. [948260]), Business Finland project Turku-R2B-Bragg WOLED and partially by the European Innovation Council through the project SCOLED (Grant Agreement Number 101098813). Views and opinions expressed are, however, those of the authors only and do not necessarily reflect those of the European Union or the European Innovation Council. Neither the European Union nor the granting authority can be held responsible for them. The authors also acknowledge Materials Research Infrastructure (MARI) at the Department of Physics and Astronomy, University of Turku, for access and support with the broad-ion beam and SEM facilities.
\end{acknowledgement}

\bibliography{references}
\newpage

\newpage
\setcounter{equation}{0}
\setcounter{figure}{0}
\setcounter{table}{0}
\setcounter{page}{1}
\makeatletter
\renewcommand{\theequation}{S\arabic{equation}}
\renewcommand{\thefigure}{S\arabic{figure}}

\begin{center}
\textbf{\Large Supporting Information}\\
\end{center}
\begin{center}
\textbf{\Large \hl{Giant Rabi splitting and polariton photoluminescence in an all solution-deposited dielectric microcavity}}\\
\end{center}
\noindent
Hassan A. Qureshi, Michael A. Papachatzakis, Ahmed Gaber Abdelmagid, Mikko Salomäki, Ermei Mäkilä, Olli Siltanen, and Konstantinos S. Daskalakis\\
\noindent
Corresponding authors: konstantinos.daskalakis@utu.fi
\section*{Contents}

\begin{figure*}[h!]
\centering
\includegraphics[width=1\linewidth]{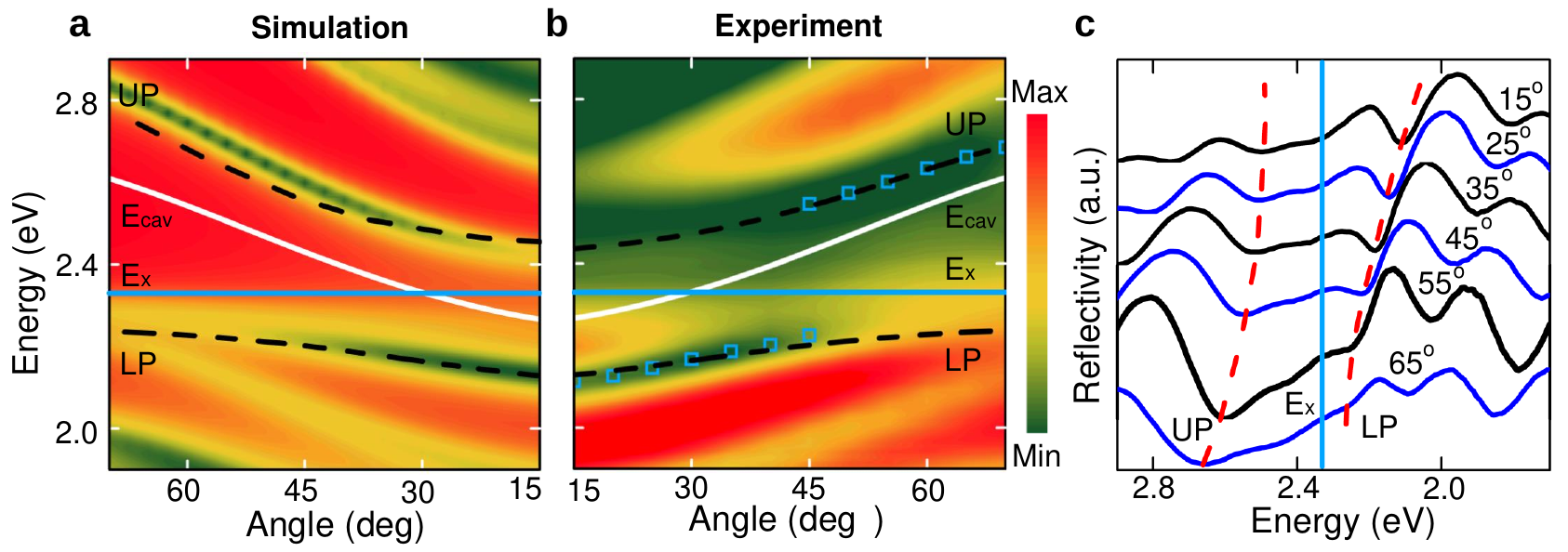}
\caption{ Reflectivity contour plots from (a) transfer matrix simulations and (b),(c) measurements of the polariton microcavity containing 3 mg/ml R6G in PVA. The individual angle-resolved reflectivity spectra are displayed on the right.}
\label{Reflectivity_Moderate_Conc}
\end{figure*}

\begin{figure*}[h!]
\centering
\includegraphics[width=0.8\linewidth]{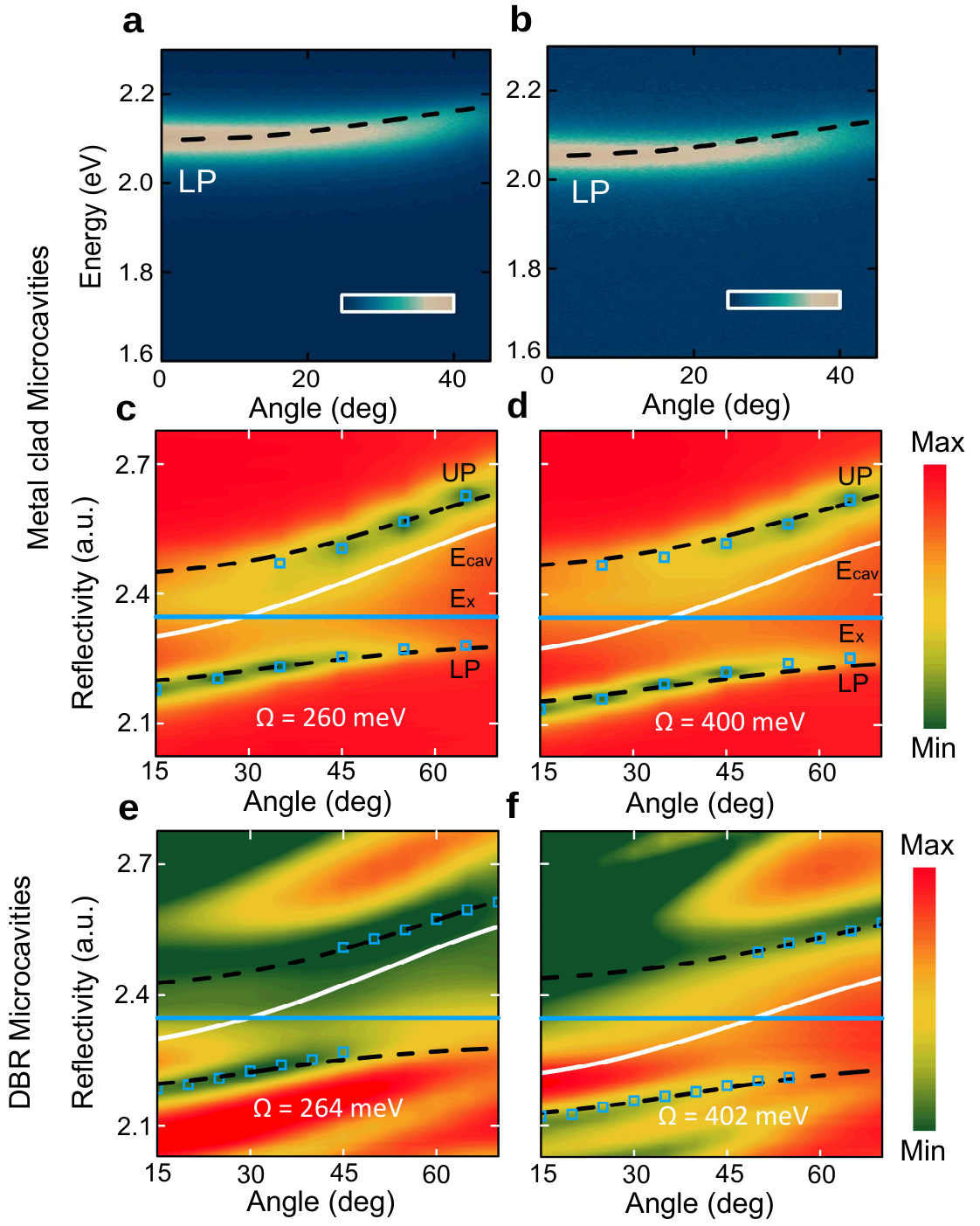}
\caption{ Photoluminescence of 3 mg/ml and 5 mg/ml silver-clad microcavities shown in (a) and (b), respectively. Photoluminescence dispersion closely matched the lower polariton dispersion calculated using a coupled harmonic oscillator model. Reflectivity of 3 mg/ml and 5 mg/ml silver-clad microcavities in (c) and (d), compared with DBR-based microcavities in (e) and (f). Blue squares are individual deeps identified from individual angle-resolved reflectivity spectra. The dashed and solid lines are fits from a coupled harmonic oscillator model; black dashed lines are polariton dispersion and solid nines are the uncoupled exciton and photon resonances.}
\label{meta_clad}
\end{figure*}

\begin{figure*}[h!]
\centering
\includegraphics[width=0.7\linewidth]{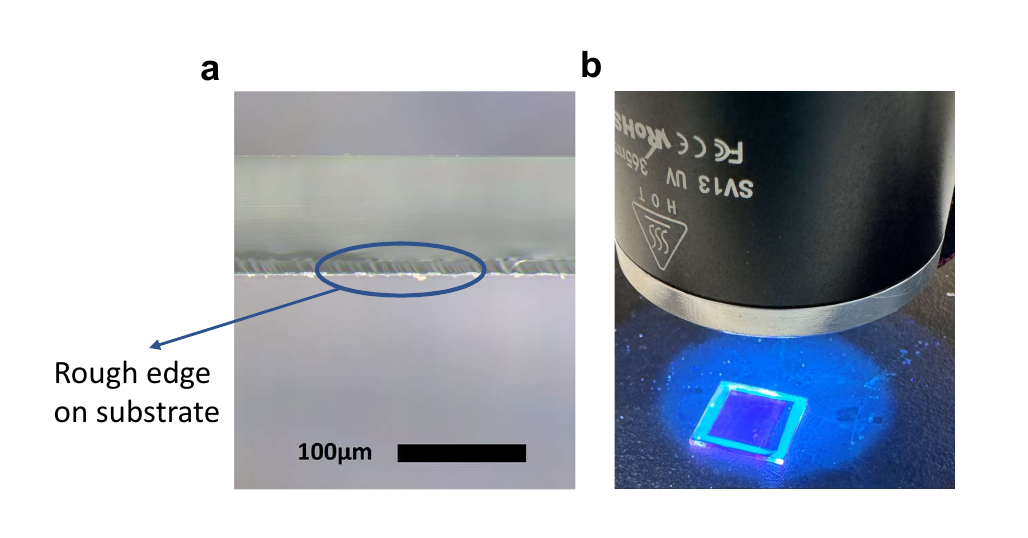}
\caption{(a) Microscope image of the glass substrate teeth-like edges. (b) Picture of half microcavity under the UV torch after applying the UV-encapsulation.}
\label{UV_glue_and_Substrate}
\end{figure*}

\begin{figure*}[h!]
\centering
\includegraphics[width=0.6\linewidth]{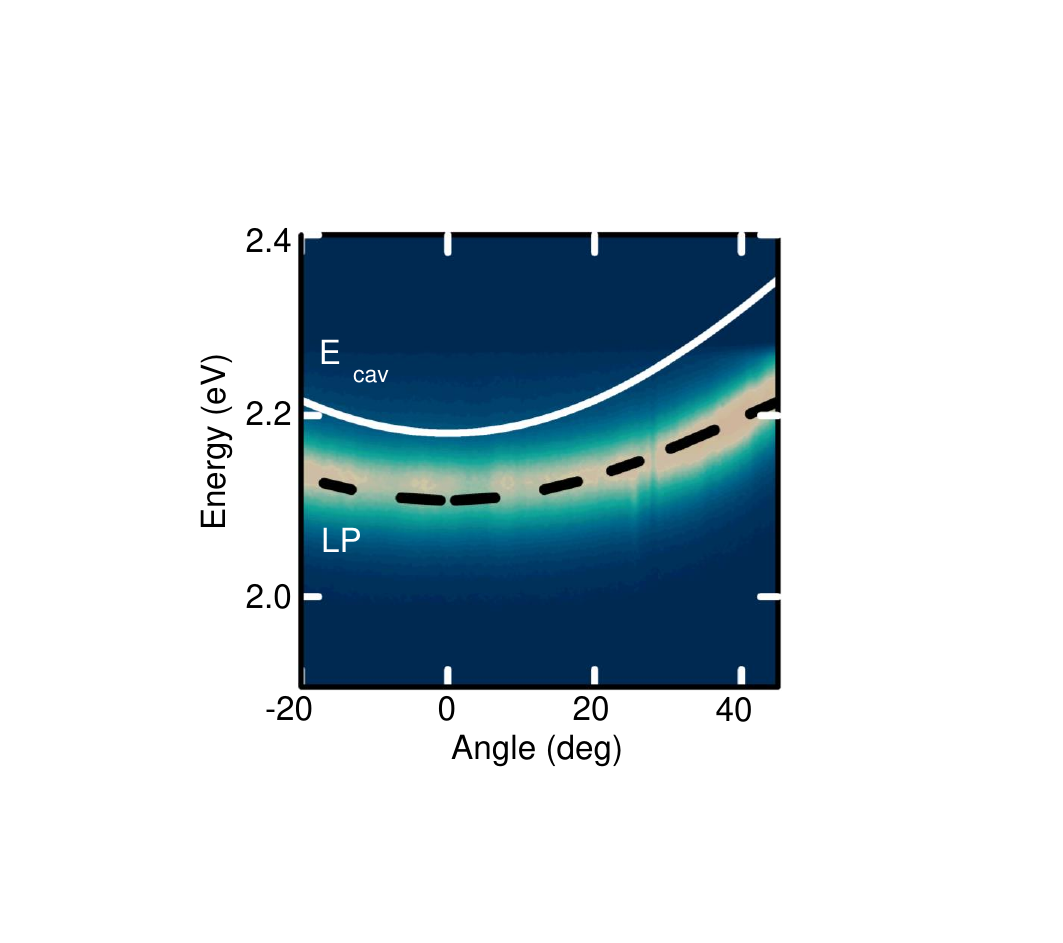}
\caption{Photoluminescence from the 3 mg/ml concentration DBR-based microcavity.}
\label{Moderate_polariton_PL}
\end{figure*}

\begin{figure*}[h!]
\centering
\includegraphics[width=1.01\linewidth]{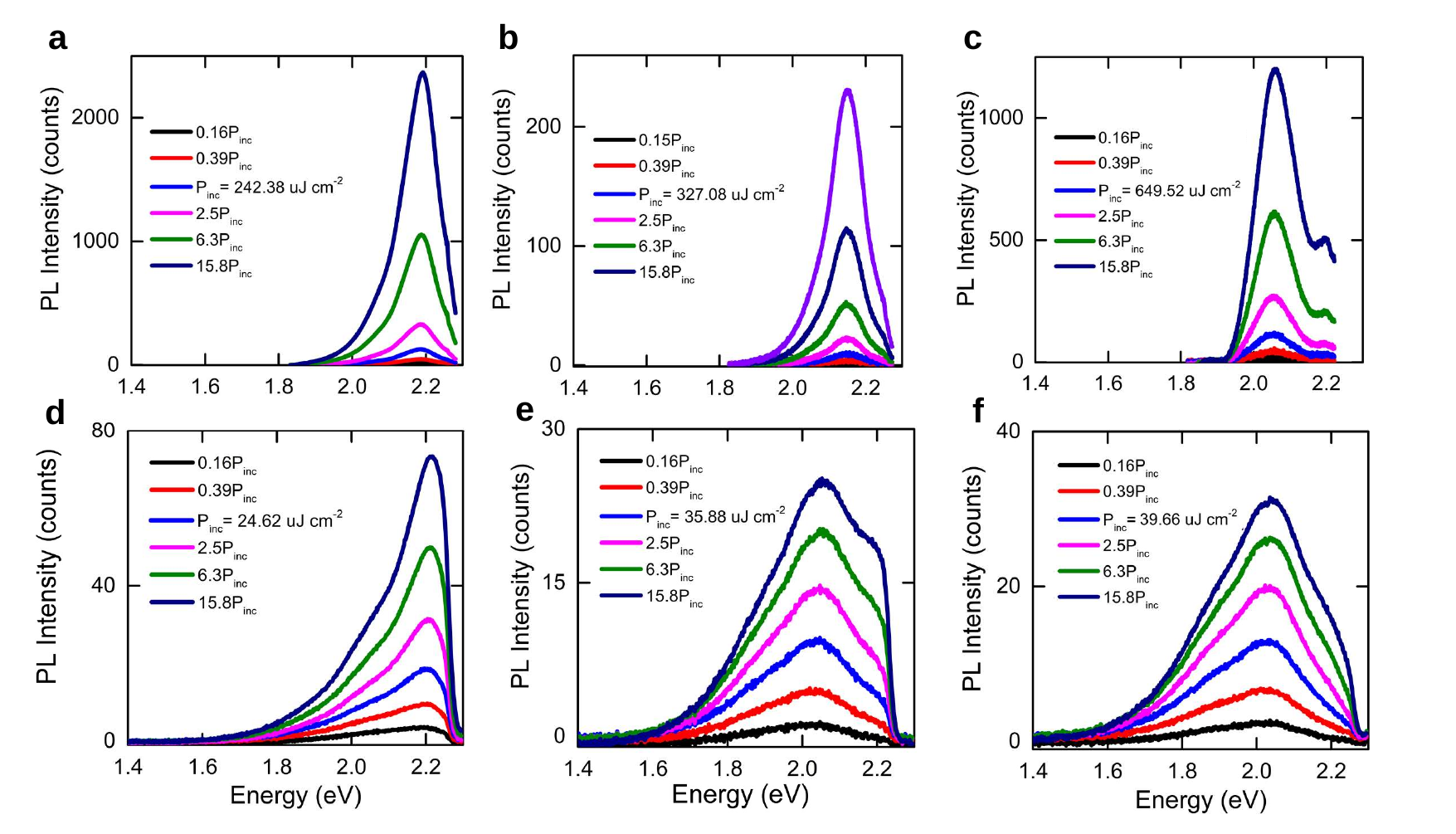}
\caption{Photoluminescence spectra recorded at different pump fluence from polariton microcavities (a)--(c) and bare films (d)--(f).}
\label{emisison_boradenning}
\end{figure*}

\begin{figure*}[h!]
\centering
\includegraphics[width=0.9\linewidth]{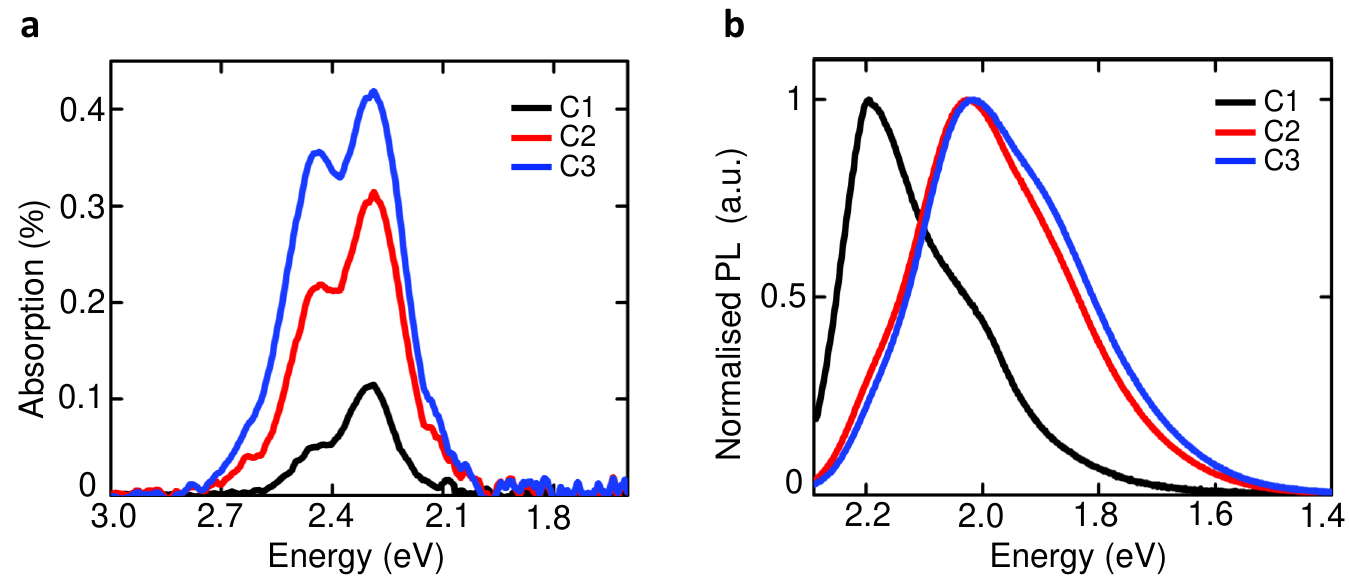}
\caption{ (a) Absorption and (b) emission spectra of 70~nm R6G in PVA films with concentrations C1 = 1~mg/ml, C2 = 3~mg/ml and C3 = 5~mg/ml. The films were deposited on glass substrates using the same spin coating recipe used for polariton microcavities. Note that we extracted absorption of the films from measured transmitted and reflected light at a 15$^\circ$ which represents the minimum angle in our setup that can measure reflectivity. }
\label{R6G}
\end{figure*}

\begin{figure*}[h!]
\centering
\includegraphics[width=0.35\linewidth]{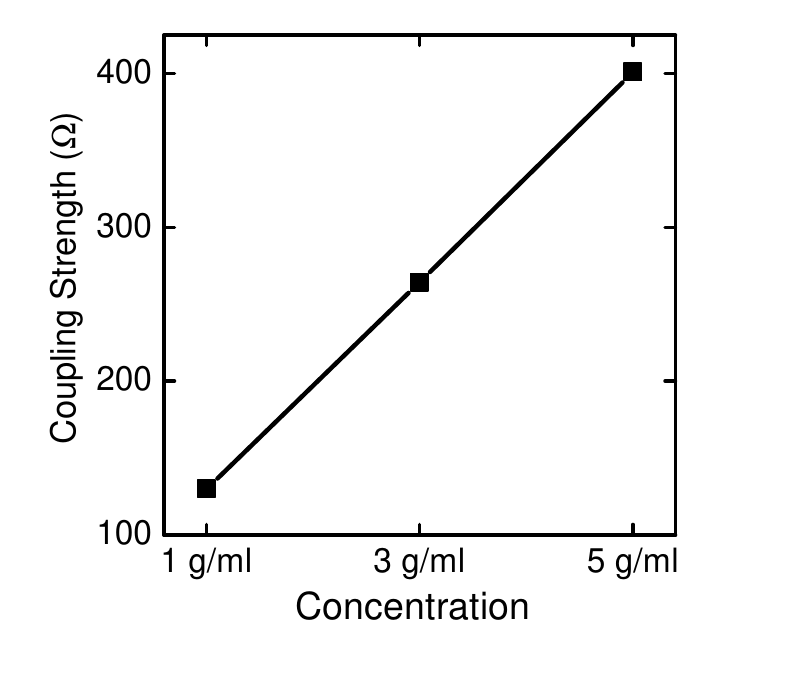}
\caption{ Rabi splitting to concentration graph in DBR-based microcavities.}
\label{Rabi vs concentration}
\end{figure*}

\begin{figure*}[h!]
\centering
\includegraphics[width=0.35\linewidth]{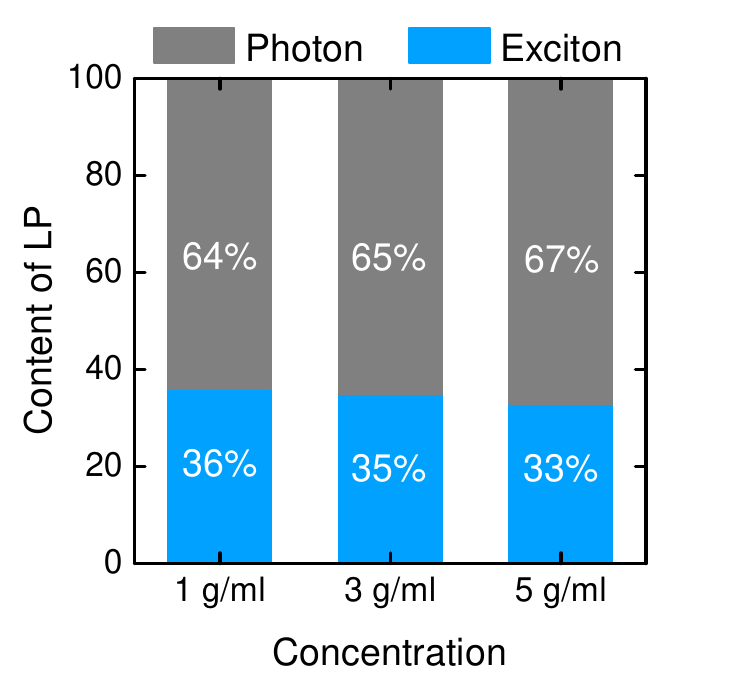}
\caption{Exciton (blue) and photon (grey) content of the LP extracted from the coupled harmonic oscillator model. 
}
\label{Hopfield}
\end{figure*}

\textbf{Supplementary Figure \ref{Reflectivity_Moderate_Conc}}. Reflectivity spectra and contour plot of polariton microcavity with 3 mg/ml R6G in PVA.\\
\textbf{Supplementary Figure \ref{meta_clad}}. Reflectivity of silver-clad microcavities and comparison with DBR-based ones.\\
\textbf{Supplementary Figure \ref{UV_glue_and_Substrate}}. Image of the substrate teeth-like edges and photo of UV-encapsulating process.\\
\textbf{Supplementary Figure \ref{Moderate_polariton_PL}}. Photoluminescence from 3 mg/ml concentration microcavity.\\
\textbf{Supplementary Figure \ref{emisison_boradenning}}. Polariton and bare film power dependent emission spectra.\\
\textbf{Supplementary Figure \ref{R6G}}. Absorption and emission of R6G.\\
\textbf{Supplementary Figure \ref{Rabi vs concentration}}. $\Omega_R$ vs concentration.\\
\textbf{Supplementary Figure \ref{Hopfield}}. Lower polariton exciton and photon content at normal collection angle.\\
\newline

\newpage

\newpage

\newpage

\newpage

\newpage

\newpage

\newpage

\newpage

\newpage

\end{document}